\documentstyle[aps,preprint]{revtex}
\begin{document}
\draft{}

\title{\hspace*{2in}IM RAS SB NNA 3-96\\[0.5cm]
 Production of scalar $K\bar K$ molecules in $\phi$ radiative decays
\thanks{This work was supported partly by RFBR, Grant 94-02-05 188,
and INTAS, Grant 94-3986 }}

\author{N.N. Achasov, V.V. Gubin and V.I. Shevchenko
\thanks{Institute for Theoretical and Experimental Physics,\ \ \
Moscow B-259,\ \ \ 117259,\ \ \ Russia}}
\address{Laboratory of Theoretical Physics\\
S.L. Sobolev Institute for Mathematics\\
Novosibirsk 90,\ \ \  630090,\ \ \ Russia
\thanks{E-mail: achasov@math.nsk.su}}
\date{\today}

\maketitle
\begin{abstract}

   The potentialities of the production of the scalar $K\bar K$ molecules in
the $\phi$ radiative decays are considered beyond the narrow resonance width
approximation.  It is shown that $BR(\phi\rightarrow\gamma f_0(a_0)\rightarrow
\gamma\pi\pi(\pi\eta))\approx (1\div 2)\times 10^{-5}\ ,\
BR(\phi\rightarrow\gamma (f_0+a_0)\rightarrow\gamma K^+K^-)\alt 10^{-6}$ and
$BR(\phi\rightarrow\gamma (f_0+a_0)\rightarrow\gamma K^0\bar K^0)\alt 10^{-8}$.
The mass spectra in the $\pi\pi\ ,\ \pi\eta\ ,\ K^+K^-\ ,\ K^0\bar K^0$
channels are calculated. The imaginary part of the amplitude
$\phi\rightarrow\gamma f_0(a_0)$ is calculated analytically. It is obtained
the phase of the scalar resonance production amplitude that causes the
interference patterns in the reaction $e^+e^-\rightarrow\gamma \pi^+\pi^-$ in
the $\phi$ meson mass region.
\end{abstract}


\section{ Introduction}
   The problem of the scalar $f_0(980)$ and $a_0(980)$ mesons has come into a
central problem of hadron spectroscopy up to charm. The point is that these
states posses peculiar properties in terms of the naive ($q\bar q$) model, see
e.g. the reviews \cite{achasov-84,achasov-91}. Whereas, all their
challenging properties can be  understood
\cite{achasov-84,achasov-91,achasov-1991} in the framework of the four-quark
($q^2\bar q^2$) MIT-bag model \cite {jaffe-77}. Along with the $q^2\bar q^2$
nature of the $a_0(980)$ and $f_0(980)$ mesons the possibility of they being
the $K\bar K$ molecules is discussed \cite {weinstein-90}. Furthermore,
probably, the $f_0(980)$ and $a_0(980)$ mesons are eye-witnesses of confinement
\cite{close-93}.

    During years of time  there was established by efforts of theorists
\cite{achasov-89,isgur-93} (see also references quoted in \cite{isgur-93}) that
the research of the decays $\phi\rightarrow\gamma f_0\rightarrow\gamma\pi\pi$
and $\phi\rightarrow\gamma a_0\rightarrow\gamma\eta\pi$ could play a crucial
role in the elucidation of the nature of the scalar $f_0(980)$ and $a_0(980)$
mesons.

   At present, the investigation of the $\phi\rightarrow\gamma
f_0\rightarrow\gamma\pi\pi$  decay by detector CMD-2 is being carried out at
an upgraded $e^+e^-$-collider VEPP-2M in Novosibirsk.
Another detector, SND, aiming to study the decays under consideration
has put into the operation at the same facility. Moreover, in the immediate
future in Frascati there is expected the start of the operation of the $\phi$
factory DA$\Phi$NE which, probably, makes possible studying the scalar
$f_0(980)$ and $a_0(980)$ mesons in the exhaustive way.

   It seems clearly that the definition of theoretical predictions prior to
data securing is a natural prerequisite to clear up a mystery of the scalar
mesons.

   In all theoretical papers (see \cite{isgur-93} and references quoted there)
except for \cite{achasov-89}, where the $q^2\bar q^2$ nature of the scalar
mesons was considered, the approximation of narrow widths of the
$f_0(980)$ and $a_0(980)$ mesons was used for their visible widths are
$25-50\ MeV$. Moreover, experimenters used this approximation to give
upper limits to branching ratios of the decays $\phi\rightarrow\gamma a_0$ and
$\phi\rightarrow\gamma f_0$ \cite{particle-94}.

   But recently \cite{achasov-95} there was shown that narrow resonance
approximation in this instance is not valid and predictions of decay
branching ratios gained in narrow resonance approximation are at least two
times overstated.

   In this connection, we study the scope for production of scalar molecules
in the $\phi$ radiative decays beyond the narrow resonance width approximation.

   In Section 2 we introduce the formulae needed for our investigation and
discuss corrections for finite widths to propagators of scalar mesons.

   Section 3 is devoted to the model of the $\phi\rightarrow\gamma a_0(f_0)$
transition amplitude. In Section 3.A  we state (more precise, make clear)
the model of the production of the scalar $K\bar K$ molecules in the $\phi$
radiative decays. In Section 3.B we calculate in the analytical form the
imaginary part of the $\phi\rightarrow\gamma a_0(f_0)$ transition amplitude
that gives near 90\%  of the branching ratios  in the $\pi\pi$ and $\eta\pi$
channels. In Section 3.C the real part of the $\phi\rightarrow\gamma a_0(f_0)$
transition amplitude, dominating in the $K\bar K$ channels, is derived partly
in the analytic form partly in the integral one suitable for the simulation
of the experimental data.

   In Section 4 we give the numerical results of our analysis. We found that in
the case of the molecule nature of the $f_0(980)$ and $a_0(980)$ resonances
$BR(\phi\rightarrow\gamma f_0(a_0)\rightarrow\gamma\pi\pi(\pi\eta))\approx
(1\div 2)\times 10^{-5}\ ,\ BR(\phi\rightarrow\gamma (f_0+a_0)\rightarrow
\gamma K^+K^-)\alt 10^{-6}$ and $BR(\phi\rightarrow\gamma (f_0+a_0)\rightarrow
\gamma K^0\bar K^0)\alt 10^{-8}$. The mass spectra for the $\pi\pi\ ,\
\pi\eta\ ,\ K^+K^-\ ,\ K^0\bar K^0$ are presented. The phase of the scalar
resonance production amplitude, causing the interference patterns in the
reaction $e^+e^-\rightarrow\gamma \pi^+\pi^-$ in the $\phi$ meson mass region,
is calculated.

   Section 5 reviews our results and discusses experimental perspectives.

\section{ Fundamental phenomenology of $\phi$ radiative decays into scalar
mesons}

   Let us introduce the amplitudes
\begin{equation}
M(\phi\rightarrow\gamma R;m)=g_R(m)(\vec e(\phi)\cdot \vec e(\gamma))\ , \quad
R=a_0\ ,\ f_0\ ,
\end{equation} 
where $\vec e(\phi)$ and $\vec e(\gamma)$ are polarization three-vectors of
the $\phi$-meson and the $\gamma$-quantum in the $\phi$-meson rest frame, $m$
is an invariant mass of two pseudoscalar mesons $a$ and $b$ produced in the
$R\rightarrow ab$ decay.

   In  $e^+e^-$ collisions the $\phi$-meson is produced transversely polarized
relative to the beam axis in the center mass system. That is why  the
amplitudes from Eq. (1) lead to the angular distribution
\begin{equation}
W(\theta)=\frac{3}{8}(1+\cos^2\theta)
\end{equation} 
if one is not interested in a polarization of the $\gamma$-quantum in the
reaction $e^+e^-\rightarrow\phi\rightarrow\gamma R$. In Eq. (2) $\theta$ is
the angle between the momentum of the $\gamma$-quantum and the axis of the
beams.

   According to the gauge invariance condition, the decay amplitudes are
proportional to the electromagnetic (electric) field, that is the photon energy
($\omega$) in a low energy region:
\begin{equation}
g_R(m)\rightarrow\omega\times const\ ,
\end{equation} 
if $m\rightarrow m_{\phi}, \> \omega=(1/2)m_{\phi}(1-m^2/m_{\phi}^2)
\rightarrow 0.$

   The decay width in narrow scalar resonance width approximation
\begin{equation}
\Gamma(\phi\rightarrow\gamma R;m)=\frac{1}{3}\frac{|g_R(m)|^2}{4\pi}
\frac{1}{2m_{\phi}}(1-\frac{m^2}{m_{\phi}^2})\ .
\end{equation} 

   The physically measurable partial widths
\begin{equation}
\Gamma(\phi\rightarrow\gamma R\rightarrow\gamma ab)=\frac{2}{\pi}
\int_{m_a+m_b}^{m_{\phi}}mdm\frac{m\Gamma(R\rightarrow ab;m)\Gamma(\phi
\rightarrow\gamma R;m)}{|D_R(m)|^2}
\end{equation} 
for $ab=\pi\pi,\ \pi^0\eta$.

   For $ab=K^+K^-,\ K^0\bar K^0$
\begin{eqnarray}
& & \Gamma(\phi\rightarrow\gamma(a_0+f_0)\rightarrow\gamma K^+K^-)=\nonumber\\
& &\frac{2}{\pi}\int_{2m_{K^+}}^{m_{\phi}}m^2\Gamma(f_0\rightarrow K^+K^-;m)
\Gamma(\phi\rightarrow\gamma f_0;m)\left|\frac{1}{D_{f_0}(m)}+
\frac{g_{a_0}(m)g_{a_0K^+K^-}}{g_{f_0}(m)g_{f_0K^+K^-}}\frac{1}{D_{a_0}(m)}
\right|^2 dm\ , \nonumber\\
& &\Gamma(\phi\rightarrow\gamma(a_0+f_0)\rightarrow\gamma K^0\bar K^0)=
\nonumber\\
& & \frac{2}{\pi}\int_{2m_{K^0}}^{m_{\phi}}m^2\Gamma(f_0\rightarrow
K^0\bar K^0;m)\Gamma(\phi\rightarrow\gamma f_0;m)\left|\frac{1}{D_{f_0}(m)}+
\frac{g_{a_0}(m)g_{a_0K^0\bar K^0}}{g_{f_0}(m)g_{f_0K^0\bar K^0}}
\frac{1}{D_{a_0}(m)}\right|^2dm\ ,
\end{eqnarray} 
where $1/D_R(m)$ is a propagator of a scalar meson.

   The width of the decay of the scalar meson $R$ into two pseudoscalar meson
state $ab$ with an invariant mass $m$ is
\begin{eqnarray}
& & \Gamma(R\rightarrow ab; m) = \frac{g^2_{Rab}}{16\pi}\frac{1}{m}
\rho_{ab}(m)\ ,\nonumber\\
& & \rho_{ab}(m) = \sqrt{(1-m^2_+/m^2)(1-m^2_-/m^2)}\enskip ,
\quad m_{\pm}= m_a\pm m_b\ .
\end{eqnarray} 
The final particle identity in the $\pi^0\pi^0$ case is taken into account in
the determination of $g_{f_0\pi^0\pi^0}$.

   In Eq. (6) it is well to bear in mind isotopic symmetry
\begin{equation}
g_{f_0K^+K^-}=g_{f_0K^0\bar K^0}\ ,\quad g_{a_0K^+K^-}=-g_{a_0K^0\bar K^0}\ .
\end{equation} 

   The use of narrow scalar resonance width approximation in the case under
consideration is not valid for two reasons\cite{achasov-95}.

   The first and major reason is connected with soft (according strong
interaction standards) photons.  From Eqs. (3), (4), (5) and (6) it follows
that the right slope of the resonance is suppressed at least by the factor
$(\omega/\omega_0)^3$, where $\quad\omega_0=m_\phi(1-M_R^2/m_{\phi}^2)/2$ and
$M_R$ is the resonance mass. As was shown \cite{achasov-95} it leads to the
suppressions of the integral contribution from the right slope of the
resonance at least by a factor of five for the decays into $\pi\pi\ ,\
\pi\eta$ channels and at least by a factor of fifty for the decays into the
$K^+K^-\ ,\ K^0\bar K^0$ channels. So, the physically measurable widths
Eq. (5) are caused practically fully by "a half " of resonance, that is by
its left slope in the channels $\pi\pi$ and $\pi\eta$. Only for this reason
the results ($\Gamma (\phi\rightarrow\gamma R; M_R)$) obtained in narrow width
approximation are overstated two times.

   The second reason is connected with finite width correction in propagators
of the scalar mesons. Let us take the generally accepted Breit - Wigner
formulae.

   If $m>2m_{K^+}\ ,\enskip 2m_{K^0}$\ ,
\begin{eqnarray}
& & \frac{1}{D_R(m)}=\frac{1}{m_R^2-m^2-i(\Gamma_0(m)+\Gamma_{K\bar K}(m))m}
\nonumber\\
& &\Gamma_{K\bar K}(m)=\frac{g^2_{RK^+K^-}}{16\pi}(\sqrt{1-4m^2_{K^+}/m^2}+
\sqrt{1-4m^2_{K^0}/m^2})\frac{1}{m}\ .
\end{eqnarray} 
If $2m_{K^+} < m < 2m_{K^0}$\ ,
\begin{equation}
\frac{1}{D_R(m)}=\frac{1}{m_R^2-m^2+ \frac{g^2_{RK^+K^-}}{16\pi}
\sqrt{4m^2_{K^0}/m^2-1} - i\frac{g^2_{RK^+K^-}}{16\pi}\sqrt{1-4m^2_{K^+}/m^2}-
i\Gamma_0(m)m}\ .
\end{equation} 
When $2m_{K^+}\ ,\enskip 2m_{K^0}>m$\ ,
\begin{equation}
\frac{1}{D_R(m)}=\frac{1}{m_R^2-m^2+ \frac{g^2_{RK^+K^-}}{16\pi}
(\sqrt{4m^2_{K^+}/m^2-1}+ \sqrt{4m^2_{K^0}/m^2-1})-
i\Gamma_0(m)m}\ .
\end{equation} 
Where the width of the decay of the scalar $R$ resonance into the $\pi\eta$ or
$\pi\pi$ channels $\Gamma_0(m)$ is determined with Eq. (7).

   Since the scalar resonances lie under the $K\bar K$ thresholds the position
of the peak in the cross section or in the mass spectrum does not coincide with
$m_R$ as is easy to see using Eqs. (9)--(11). That is why it is necessary to
renormalize the mass in the  Breit - Wigner formulae, Eqs. (9)--(11),
\begin{eqnarray}
m^2_R = M^2_R - \frac{g^2_{RK^+K^-}}{16\pi}(\sqrt{4m^2_{K^+}/M_R^2-1}+
\sqrt{4m^2_{K^0}/M_R^2-1})\ ,
\end{eqnarray} 
where $M^2_R$ is the physical mass squared ( $M_{a_0} = 980\ MeV$ and $M_{f_0}
= 980\ MeV$ ) while $m^2_R$ is the bare mass squared. So, the physical mass is
heavier than the bare one. This circumstance is especially important in the
case of a strong coupling of the scalar mesons with the $K\bar K$ channels as
in the four-quark and molecule models. Meanwhile it was not taken into account
both in fitting of data and theoretical works excluding the papers
\cite{achasov-84,achasov-89,achasov-88}. It should be noted that Eqs. (9)--(11)
are applied only in a resonance region. They, for example, have incorrect
analytic properties at $m^2=0$. The expressions, in which this defect is
removed, can be found in \cite{achasov-84,achasov-89,achasov-88}.

   Notice that when the scalar resonances lie between the $K^+K^-$ and
$K^0\bar K^0$ thresholds it is necessary to renormalize the mass in the
Breit - Wigner formulae in the following manner
\begin{eqnarray}
m^2_R = M^2_R - \frac{g^2_{RK^+K^-}}{16\pi}\sqrt{4m^2_{K^0}/M_R^2-1}\ .
\end{eqnarray} 

   The coupling constants in the molecule model
\cite{weinstein-90,isgur-93,close-92}:
\begin{equation}
g^2_{f_0K^+K^-}/4\pi=g^2_{a_0K^+K^-}/4\pi=0.6\ GeV^2\ .
\end{equation} 
Notice that in this model $M_R-m_R=24(10)\ MeV$ for $M_R=980(2m_{K^+})\ MeV$.

\section{ Model of $K\bar K$ scalar molecule production in $\phi$ radiative
decays}
\subsection{$K\bar K$ loop production of extended scalar mesons}

   Unfortunately, at present it is not possible to construct the truly
relativistic gauge invariant model in the case of a $K\bar K$ scalar molecule
production in the $\phi$ radiative decays for the nonrelativistic nature of
a wave function of a molecule. But it is possible to "relativize" a model
constructed in the $K\bar K$ molecule rest frame and to place a gauge
invariance constraint.

   In the rest frame of a scalar $K\bar K$ molecule we consider the production
mechanism described by the diagrams in Fig.1, where the cross in the
$K^+K^-f_0(a_0)$ vertex points to a coupling of the $K^+K^-$ state with  a
extended meson.

   In the scalar molecule rest frame, $\vec p=\vec q$, the decay amplitude
\begin{eqnarray}
& & T(p,q) = M(p,q) - M(p,0)\nonumber\ ,\\
& & M(p,q) = M_1(p,q) + M_2(p,q) + M_3(p,q)\ ,
\end{eqnarray} 
where $p$ and $q$ are four-momenta of the $\phi$ meson and the photon
respectively. The amplitudes
\begin{eqnarray}
& & M_1(p,q)=ieg_{RK^+K^-}g_{\phi K^+K^-}\epsilon^{\mu}(\phi)\epsilon^{\nu}(
\gamma)\int\frac{d^4k}{(2\pi)^4}\phi(|\vec k|)\frac{2g_{\mu\nu}}{D(k)D(k-p+q)}
\ ,\\
& & M_2(p,q)=\nonumber\\
& & -ieg_{RK^+K^-}g_{\phi K^+K^-}\epsilon^{\mu}(\phi)\epsilon^{\nu}(\gamma)
\int\frac{d^4k}{(2\pi)^4}\phi(|\vec k|)\frac{(2k+q)_{\nu}(2k+2q-p)_{\mu}}
{D(k)D(k-p+q)D(k+q)}\ , \\ 
& & M_3(p,q)=\nonumber\\
& & -ieg_{RK^+K^-}g_{\phi K^+K^-}\epsilon^{\mu}(\phi)\epsilon^{\nu}(\gamma)
\int\frac{d^4k}{(2\pi)^4}\phi(|\vec k|)\frac{(2k-2p+q)_{\nu}(2k-p)_{\mu}}
{D(k)D(k-p+q)D(k-p)}  
\end{eqnarray}
correspond to the diagrams Fig.1a, Fig.1b, Fig.1c respectively. The function
$\phi(|\vec k|)$ describes a momentum distribution of a $K^+(K^-)$
meson in a molecule, $D(k)=k^2-m^2+i0$, $\epsilon^{\mu}(\phi)$ and
$\epsilon^{\nu}(\gamma)$ are polarization four-vectors of the $\phi$ meson and
the $\gamma$-quantum, $e^2/4\pi=\alpha\simeq 1/137$, $g_{RK^+K^-}$ is
determined by Eq. (7) and $g_{\phi K^+K^-}$ is determined by the following
manner
\begin{eqnarray}
\Gamma (\phi\rightarrow K^+K^-)=\frac{1}{3}\frac{g_{\phi K^+K^-}^2}{16\pi}
m_{\phi}\left(\sqrt{1-\frac{4m_{K^+}^2}{m_{\phi}^2}}\right)^3\ .
\end{eqnarray} 

   The subtraction at $q=0$ is gauge invariant regularization that always is
necessary in the field theory.

   The "relativization"
\begin{eqnarray}
T(p,q)=(\vec e(\phi)\cdot\vec e(\gamma))g_R(m)\equiv g_R(m)[(e(\phi)\cdot
e(\gamma))-(e(\phi)q)(e(\gamma)p)/(pq)]|_{\vec p=\vec q}\ .
\end{eqnarray} 
gives the relativistic and gauge invariant amplitude. Emphasize that in
Eq. (20)  $\vec e(\phi)$ and $\vec e(\gamma)$ are polarization three-vectors
of the $\phi$-meson and the $\gamma$-quantum in the molecule rest frame.

   When $\phi (|\vec k|)=1$ Eqs. (15)--(18) reproduce the well definite gauge
invariant field theory expression \cite{achasov-89}.

   Our model is the practically same one as in \cite{isgur-93}. Nevertheless,
it should be noted that in Eqs. (4.21)--(4.23) of the paper \cite{isgur-93}
the function $\phi(|\vec k|)$ is used to describe a momentum
distribution of a $K^+(K^-)$ meson in a molecule whereas a $K^+(K^-)$ meson
has the three-momentum equal $\pm (\vec k - \vec q/2)$. So, Eqs. (4.21)--(4.23)
of \cite{isgur-93} are not quite correct and our Eqs. (15)--(18) make clear the
 molecule production model
\footnote{In contrast to \cite{isgur-93} we treat with an amplitude rather
than a S-matrix element. The difference is the factor $i$.

   Besides, note that in Eq. (4.24) of \cite{isgur-93} the sign (-) is lost!
Probably, it is a misprint otherwise the gauge invariance constraint, see Eq.
(3), is destroyed.}.

   Let us integrate Eqs. (16)--(18) over $k_0$ in the molecule rest frame.
\begin{eqnarray}
&&M_1(p;q)=-2eg_{RK^+K^-}g_{\phi K^+K^-}\vec\epsilon(\phi)\vec\epsilon(\gamma)
\int\frac{d^3k}{(2\pi)^3}\phi(|\vec k|)\frac{1}{E_k(m^2-4E_k^2)}\ , \\
&&M_2(p;q)=eg_{RK^+K^-}g_{\phi K^+K^-}\vec\epsilon(\phi)\vec\epsilon(\gamma)
\int\frac{d^3k}{(2\pi)^3}\phi(|\vec k|)\left(\vec k^2-\frac{(\vec k\vec q)^2}
{\vec q^2}\right)\times\nonumber\\
&&\times\Biggl(\frac{1}{E_k(m^2+2E_km)(2\omega E_k+2\vec k\vec q)}+
\frac{1}{E_k(m^2-2E_km)(2p_0E_k-m_{\phi}^2+2\vec k\vec q)}-\nonumber\\
&&-\frac{1}{E_{k+q}(2\omega^2+2\omega E_{k+q}+2\vec k \vec q)(p_0^2+
\omega^2+2p_0E_{k+q}+2\vec k \vec q)}\Biggr)\ , \\ 
&&M_3(p;q)=-eg_{RK^+K^-}g_{\phi K^+K^-}\vec \epsilon(\phi)\vec \epsilon(\gamma)
\int\frac{d^3k}{(2\pi)^3}\phi(|\vec k|)\left(\vec k^2-
\frac{(\vec k\vec p)^2}{\vec p^2}\right)\times\nonumber\\
&&\times\Biggl(\frac{1}{E_k(m^2+2E_km)(2p_0E_k+m_{\phi}^2+2\vec k\vec p)}+
\frac{1}{E_k(m^2-2E_km)(2\omega E_k+2\vec k\vec p)}+\nonumber\\
&&+\frac{1}{E_{k-q}(\omega^2+\vec p^2-2\omega E_{k-q}-2\vec k \vec p)
(p_0^2+\vec p^2-2p_0E_{k-q}-2\vec k \vec p)}\Biggr)\ , 
\end{eqnarray}
where
\begin{eqnarray}
&&\vec p=\vec q\ ,\ p_0-q_0=m\ ,\ q_0=|\vec q|=\omega=(m_{\phi}^2-m^2)/2m\ ,\
E_k=\sqrt{\vec k^2+m_{K^+}^2-i0}\ ,\nonumber\\
&&E_{k+q}=\sqrt{(\vec k+\vec q)^2+m_{K^+}^2-i0}\ ,\ E_{k-q}=\sqrt{(\vec k-
\vec q)^2+m_{K^+}^2-i0}\ .
\end{eqnarray} 

   It is easily seen that the integration over the angle between $\vec k$ and
$\vec q$ (or $\vec p$) is performed analytically.

\subsection{Imaginary part of amplitude of $\phi\rightarrow\gamma f_0(a_0)$
decay}

   Following \cite {isgur-93,close-92} we use
\begin{equation}
\phi(|\vec k|)=\frac{\mu^4}{(|\vec k|^2+\mu^2)^2}
\end{equation}  
where $\mu =141\ MeV$.

   The function $\phi(|\vec k|)$ suppresses the contribution of virtual
$K^+K^-$ pairs . That is why it is natural to expect that the imaginary part
of the $\phi\rightarrow\gamma f_0(a_0)$ amplitude is essential in comparison
with the real one. The numerical analyses in Section 4 supports this hope.

   So, the imaginary part of the $\phi\rightarrow\gamma f_0(a_0)$ amplitude
\begin{eqnarray}
&&ImT(p,q)=Im(M(p,q)-M(p,0))=\vec\epsilon(\phi)\vec\epsilon(\gamma)Img_R(m)=
\nonumber\\
&&=\vec\epsilon(\phi)\vec\epsilon(\gamma)Im(\bar g_R(m) - \bar g_R(m_{\phi}))
\end{eqnarray}  
merits the individual consideration here. Eq. (25) makes possible calculating
the imaginary part in the analytical form. This calculation is too cumbersome
to present it here. Because of this we shall only explain the genesis of the
imaginary part of the amplitude and shall adduce its analytical form.

   The contributors of the imaginary part of the $\phi\rightarrow\gamma
f_0(a_0)$ amplitude are the regions of the integration where the denominators in
Eqs. (21)--(23) vanish. We treat these zeros using an infinitesimal negative
imaginary addition to the $K^+$ mass square ($m_{K^+}^2-i0$), see Eq. (24).

   When  $m<2m_{K^+}$
\begin{eqnarray}
&&Img_R(m)=\pi eg_{RK^+K^-}g_{\phi K^+K^-}\frac{\mu^4}{(2\pi)^2}\frac{1}
{(m^2-4a^2)^2}\Biggl\{\frac{m_{\phi}^2}{\omega^3}\Biggl(\ln\frac{(E_1-a)
(E_2+a)}{(E_2-a)(E_1+a)}\times\nonumber\\
&\times&\frac{E_1E_2m^2(12a^2-m^2)-a^2m_{\phi}^2(m^2+4a^2)}{4a^3m^2}\Biggr)+
\nonumber\\
&+&\frac{4m_{K^+}^2}{m\omega}\ln\frac{E_1^2-a^2}{E_2^2-a^2}-\frac{8m_{K^+}^2}{m
\omega}\ln\frac{m_{\phi}-\sqrt{m_{\phi}^2-4m_{K^+}^2}}{m_{\phi}+
\sqrt{m_{\phi}^2-4m_{K^+}^2}}+\nonumber\\
&+&\frac{m_{\phi}^2(m^2-4a^2)(E_1-E_2)}{2a^2\omega^3}-
\frac{32(m_{\phi}^2-4m_{K^+}^2)^{3/2}(m^2-4a^2)^2}{3m_{\phi}(m_{\phi}^2-
4a^2)^3}\Biggr\}
\end{eqnarray} 
where $a^2=m_{K^+}^2-\mu^2$, $E_1=\frac{p_0}{2}-
\frac{\omega}{2m_{\phi}}\sqrt{m_{\phi}^2-4m_{K^+}^2}$\ , and
$E_2=\frac{p_0}{2}+\frac{\omega}{2m_{\phi}}\sqrt{m_{\phi}^2-4m_{K^+}^2}$\ .

   When $m>2m_{K^+}$
\begin{eqnarray}
&&Img_R(m)=\pi eg_{RK^+K^-}g_{\phi K^+K^-}\frac{\mu^4}{(2\pi)^2}\frac{1}{(m^2-
4a^2)^2}\Biggl\{\frac{m_{\phi}^2}{\omega^3}\Biggl(\ln\frac{(E_1-a)(E_2+a)}{(E_2
-a)(E_1+a)}\times\nonumber\\
&\times&\frac{E_1E_2m^2(12a^2-m^2)-a^2m_{\phi}^2(m^2+4a^2)}{4a^3m^2}\Biggr)+
\nonumber\\
&+&\frac{4m_{K^+}^2}{m\omega}\ln\frac{E_1^2-a^2}{E_2^2-a^2}-\frac{8m_{K^+}^2}{m
\omega}\ln\frac{m_{\phi}-\sqrt{m_{\phi}^2-4m_{K^+}^2}}{m_{\phi}+
\sqrt{m_{\phi}^2-4m_{K^+}^2}}+\nonumber\\
&+&\frac{m_{\phi}^2(m^2-4a^2)(E_1-E_2)}{2a^2\omega^3}-
\frac{32(m_{\phi}^2-4m_{K^+}^2)^{3/2}(m^2-4a^2)^2}{3m_{\phi}(m_{\phi}^2-
4a^2)^3}+\nonumber\\
&+&\frac{4\sqrt{m^2-4m_{K^+}^2}m_{\phi}^2}{\omega m^2}+\frac{8m_{K^+}^2}{\omega
m}\ln\frac{m-\sqrt{m^2-4m_{K^+}^2}}{m+\sqrt{m^2-4m_{K^+}^2}}\Biggr\}
\end{eqnarray} 

   Notice that
\begin{equation}
Im\bar g_R(m_{\phi})=eg_{RK^+K^-}g_{\phi K^+K^-}\frac{8}{3\pi}
\frac{\mu^4(m_{\phi}^2-4m^2_{K^+})}{(m_{\phi}^2-4a^2)^3}\sqrt{1-
\frac{4m_{K^+}^2}{m_{\phi}^2}}\ .
\end{equation} 

\subsection{Real part of amplitude of $\phi\rightarrow\gamma f_0(a_0)$
decay}

   As will be seen from Section 4 the real part of $\phi\rightarrow\gamma
f_0(a_0)$ amplitude
\begin{eqnarray}
&&ReT(p,q)=Re(M(p,q)-M(p,0))=\vec\epsilon(\phi)\vec\epsilon(\gamma)Reg_R(m)=
\nonumber\\
&&=\vec\epsilon(\phi)\vec\epsilon(\gamma)Re(\bar g_R(m) - \bar g_R(m_{\phi}))
\end{eqnarray}  
dominates in the suppressed $K^+K^-$ and $K^0\bar K^0$ channels.

   When $m<2m_{K^+}$
\begin{eqnarray}
&&Reg_R(m)=eg_{RK^+K^-}g_{\phi K^+K^-}\frac{\mu^4}{(2\pi)^2}\times\nonumber\\
&&\times\Biggl\{\frac{8m_{\phi}^2}{\omega m(m^2-4a^2)^2}[\frac{\mu}{a}
\arctan\frac{a}{\mu}-\frac{\sqrt{4m_{K^+}^2-m^2}}{m}\arctan\frac{m}{
\sqrt{4m_{K^+}^2-m^2}}]+\nonumber\\
&&+\frac{8m_{K^+}^2}{\omega m(m^2-4a^2)^2}[(\arctan\frac{a}{\mu})^2-
(\arctan\frac{m}{\sqrt{4m_{K^+}^2-m^2}})^2]+\nonumber\\
&&\frac{1}{(m^2-4a^2)a^2}(2+\frac{m}{2\omega}+\frac{p_0(m^2-4a^2)}{2\omega^2m})
(1-\frac{m_{K^+}^2}{\mu a}\arctan\frac{a}{\mu})+\frac{2}{\omega m(m^2-4a^2)}+
\nonumber\\
&&+\int_0^{\infty}\frac{|\vec k|d|\vec k|}{(\vec k^2+\mu^2)^22\omega^3}
\Biggl(E_+-E_--\nonumber\\
&&-\frac{\omega^2m_{K^+}^2}{E_k(2E_km-m^2)}\ln\frac{(E_k+\omega+E_-)(E_k+
\omega-E_+)}{(E_k+\omega+E_+)(E_k+\omega-E_-)}-\nonumber\\
&&-\frac{\omega^2m_{K^+}^2}{E_km(2E_k+m)}\ln\frac{(E_--E_k+\omega)(E_k-\omega+
E_+)}{(E_+-E_k+\omega)(E_-+E_k-\omega)}+\nonumber\\
&&+\frac{m_{\phi}^2(E_k+E_1)(E_k+E_2)}{E_km(2E_k+m)}\ln\frac{(E_k+p_0+E_-)^2}
{(E_k+p_0+E_+)^2}+\nonumber\\
&&+\frac{m_{\phi}^2(E_k-E_1)(E_k-E_2)}{E_km(2E_k-m)}\ln
\frac{(E_k-p_0+E_+)^2}{(E_k-p_0+E_-)^2}\Biggr)\Biggr\}-Re\bar g_R(m_{\phi})
\end{eqnarray}  
where $E_+=\sqrt{\vec k^2 + \omega^2 + 2|\vec k|\omega + m_{K^+}^2}$\ ,\
$E_-=\sqrt{\vec k^2 + \omega^2 - 2|\vec k|\omega + m_{K^+}^2}$ and
\begin{eqnarray}
&&Re\bar g_R(m_{\phi})=eg_{RK^+K^-}g_{\phi K^+K^-}\frac{\mu^4}{(2\pi)^2}
\Biggl\{[\frac{\mu}{a}\arctan\frac{a}{\mu}-\nonumber\\
&&-\frac{\sqrt{m_{\phi}^2-4m_{K^+}^2}}{2m_{\phi}}\ln\frac{m_{\phi}+
\sqrt{m_{\phi}^2-4m_{K^+}^2}}{m_{\phi}-\sqrt{m_{\phi}^2-4m_{K^+}^2}}](\frac{64
(m_{\phi}^4-3m_{\phi}^2m_{K^+}^2-4m_{K^+}^2a^2)}{3m_{\phi}^2(m_{\phi}^2-
4a^2)^3})-\nonumber\\
&-&\arctan(\frac{a}{\mu})(\frac{2m_{K^+}^2}{\mu a^3(m_{\phi}^2-4a^2)}+
\frac{8\mu m_{K^+}^2}{3a^3m_{\phi}^2(m_{\phi}^2-4a^2)}-\frac{16m_{K^+}^2\mu}
{3a^3(m_{\phi}^2-4a^2)^2}+\nonumber\\
&&+\frac{2m_{K^+}^2\mu}{m_{\phi}^2a^5})+\frac{2(5m_{\phi}^2m_{K^+}^2-
20m_{K^+}^4-2m_{\phi}^2\mu^2+12m_{K^+}^2\mu^2+8\mu^4)}
{3a^4(m_{\phi}^2-4a^2)^2}+\nonumber\\
&&+\frac{32m_{K^+}^2\sqrt{m_{\phi}^2-4m_{K^+}^2}}
{3m_{\phi}^3(m_{\phi}^2-4a^2)^2}\ln\frac{m_{\phi}+
\sqrt{m_{\phi}^2-4m_{K^+}^2}}{m_{\phi}-\sqrt{m_{\phi}^2-4m_{K^+}^2}}\Biggr\}\ .
\end{eqnarray}  

   When $m>2m_{K^+}$
\begin{eqnarray}
&&Reg_R(m)=eg_{RK^+K^-}g_{\phi K^+K^-}\frac{\mu^4}{(2\pi)^2}\times\nonumber\\
&&\times\Biggl\{\frac{8m_{\phi}^2}{\omega m(m^2-4a^2)^2}[\frac{\mu}{a}
\arctan\frac{a}{\mu}-\frac{\sqrt{m^2-4m_{K^+}^2}}{2m}\ln\frac{m+\sqrt{m^2-
4m_{K^+}^2}}{m-\sqrt{m^2-4m_{K^+}^2}}]+\nonumber\\
&&+\frac{8m_{K^+}^2}{\omega m(m^2-4a^2)^2}[(\arctan\frac{a}{\mu})^2+
\frac{1}{4}\ln^2\frac{m+\sqrt{m^2-4m_{K^+}^2}}{m-\sqrt{m^2-4m_{K^+}^2}}-
\frac{\pi^2}{4}]+\nonumber\\
&&+\frac{1}{(m^2-4a^2)a^2}(2+\frac{m}{2\omega}+\frac{p_0(m^2-4a^2)}
{2\omega^2m})(1-\frac{m_{K^+}^2}{\mu a}\arctan\frac{a}{\mu})+\nonumber\\
&+&\frac{2}{\omega m(m^2-4a^2)}-\frac{m_{K^+}^2}{2\omega m}\Biggl(\frac{4}
{(m^2-4a^2)^2}[\frac{m}{2a}
\ln\frac{m_{K^+}-a}{m_{K^+}+a}+
\ln\frac{4\mu^2}{(2m_{K^+}^2-m)^2}]-\nonumber\\
&-&\frac{1}{2(m^2-4a^2)}[\frac{m}{2a^3}\ln\frac{m_{K^+}-a}
{m_{K^+}+a}+\frac{mm_{K^+}+2a^2}{a^2\mu^2}]\Biggr)\ln\frac{m+\sqrt{m^2-
4m_{K^+}^2}}{m-\sqrt{m^2-4m_{K^+}^2}}\nonumber\\
&+&\int_0^{\infty}\frac{|\vec k|d|\vec k|}{(\vec k^2+\mu^2)^22\omega^3}
\Biggl(E_+-E_-+\frac{\omega^2m_{K^+}^2}{E_k(m^2-2E_km)}\times\nonumber\\
&&\times\ln\frac{(E_k+\omega+E_-)(E_k+
\omega-E_+)}{(E_k+\omega+E_+)(E_k+\omega-E_-)}-\nonumber\\
&-&\frac{\omega^2m_{K^+}^2}{E_km(2E_k+m)}\ln\frac{(E_--E_k+\omega)(E_k-\omega+
E_+)}{(E_+-E_k+\omega)(E_-+E_k-\omega)}+\nonumber\\
&+&\frac{m_{\phi}^2(E_k+E_1)(E_k+E_2)}{E_km(2E_k+m)}\ln\frac{(E_k+p_0+E_-)^2}
{(E_k+p_0+E_+)^2}-\nonumber\\
&-&\frac{m_{\phi}^2(E_k-E_1)(E_k-E_2)}{E_k(m^2-2E_km)}\ln
\frac{(E_k-p_0+E_+)^2}{(E_k-p_0+E_-)^2}+\nonumber\\
&+&\frac{m_{K+}^2\omega^2}{E_k(m^2-2E_km)}\ln
\frac{m-\sqrt{m^2-4m_{K^+}^2}}{m+\sqrt{m^2-4m_{K^+}^2}}\Biggr)\Biggr\}-
Re\bar g_R(m_{\phi})\ .
\end{eqnarray} 

   It is easy to verify that the integrand in Eq. (33) is nonsingular at
$2E_k=m$ for the principal value of the integral was calculated analytically.
That is why Eq. (33) is suitable for a simulation of experimental data.

   \section{ Quantitative results }

   Below we consider two variants: i) $\Gamma_0(M_R) =50\ MeV$ that
corresponds to the visible width $\simeq25\ MeV$ and the partial intensity of
the decay into the $K\bar K$ channels $BR(R\rightarrow K\bar K)\simeq0.35$ in
the molecule model, Eq. (14), and ii) $\Gamma_0(M_R)=100\ MeV$ that
corresponds to the visible width $\simeq75\ MeV$ and $BR(R\rightarrow K\bar K)
\simeq0.3$ in the molecule model, Eq. (14). Please see the definition of
$\Gamma_0(m)$ in Eqs. (9)--(11), (7).

   We present $BR(\phi\rightarrow\gamma R;m)=
\Gamma(\phi\rightarrow\gamma R;m)/\Gamma (\phi)$, where $\Gamma (\phi)$ is the
$\phi$ meson full width, in Fig.2, the phase of $-g_R(m)$ ( $\delta =\arctan
\left (Img_R(m)/Reg_R(m)\right )$ ) in Fig.3, the spectra
\begin{eqnarray}
\frac{dBR(\phi\rightarrow\gamma R\rightarrow\gamma ab)}{dm}=
\frac{d\Gamma(\phi\rightarrow\gamma R\rightarrow\gamma ab)}{\Gamma (\phi)dm}=
\frac{2}{\pi}\frac{m^2\Gamma(R\rightarrow ab;m)
\Gamma(\phi\rightarrow\gamma R;m)}{\Gamma (\phi)|D_R(m)|^2}
\end{eqnarray} 
for $ab=\pi\pi$
\footnote{ Notice that $d\Gamma(\phi\rightarrow\gamma R\rightarrow\gamma
\pi\pi)/dm=d\Gamma(\phi\rightarrow\gamma R\rightarrow\gamma
\pi^+\pi^+)/dm+d\Gamma(\phi\rightarrow\gamma R\rightarrow
\gamma \pi^0\pi^0)/dm= 1.5\cdot d\Gamma(\phi\rightarrow\gamma R
\rightarrow\gamma \pi^+\pi^+)/dm $.}
and $\pi\eta$ for the different $\Gamma_0(M_R)$ in Fig.4 and Fig.5
respectively. The spectra
\begin{eqnarray}
& &dBR(\phi\rightarrow\gamma(a_0+f_0)\rightarrow\gamma K^+K^-)/dm=
d\Gamma(\phi\rightarrow\gamma(a_0+f_0)\rightarrow\gamma K^+K^-)/
\Gamma (\phi)dm=\nonumber\\
& &=\frac{2}{\pi}m^2\Gamma(f_0\rightarrow K^+K^-;m)\frac{
\Gamma(\phi\rightarrow\gamma f_0;m)}{\Gamma (\phi)}\left|\frac{1}{D_{f_0}(m)}+
\frac{1}{D_{a_0}(m)}\right|^2dm
\end{eqnarray}  
and
\begin{eqnarray}
& &dBR(\phi\rightarrow\gamma(a_0+f_0)\rightarrow\gamma K^0\bar K^0)/dm=
d\Gamma(\phi\rightarrow\gamma(a_0+f_0)\rightarrow\gamma K^0\bar K^0)/
\Gamma (\phi)dm=\nonumber\\
& &=\frac{2}{\pi}m^2\Gamma(f_0\rightarrow K^0\bar K^0;m)\frac{
\Gamma(\phi\rightarrow\gamma f_0;m)}{\Gamma (\phi)}\left|\frac{1}{D_{f_0}(m)}-
\frac{1}{D_{a_0}(m)}\right|^2dm
\end{eqnarray} 
are presented in Fig.6 and Fig.7 respectively.

   As with the four-quark model \cite{achasov-89} in the molecule case the
interference is constructive in the $K^+K^-$ channel and destructive in the
$K^0\bar K^0$ channel, see Eqs. (35) and (36), for the $K^+K^-$ loop diagram
production model.

   We study the dependence of the branching ratios under consideration on
the resonance mass. Our results are listed at Tables I--III. Calculating
Table I we take into account only the imaginary part of the scalar resonance
production amplitudes.

\section{ Conclusion }

   At $m_{f_0}=980\ MeV$ and $\Gamma_0(m_{f_0})=50(100)\ MeV$
$BR(\phi\rightarrow\gamma f_0\rightarrow\gamma\pi\pi = 1.9(2.2)\cdot 10^{-5}$.
At $m_{a_0}=980\ MeV$ and $\Gamma_0(m_{a_0})=50(100)\ MeV$
$BR(\phi\rightarrow\gamma a_0\rightarrow\gamma\pi\eta = 1.9(2)\cdot 10^{-5}$.

   As seen from Tables $BR(\phi\rightarrow\gamma f_0(a_0)
\rightarrow\gamma\pi\pi(\pi\eta))\approx (1\div 2)\times 10^{-5}\ ,\
BR(\phi\rightarrow\gamma (f_0+a_0)\rightarrow\gamma K^+K^-)\alt 10^{-6}$ and
$BR(\phi\rightarrow\gamma (f_0+a_0)\rightarrow\gamma K^0\bar K^0)\alt 10^{-8}$.

   When the branching ratios from Table I is compared with the ones in the
narrow resonance approximation on Fig. 2, it is apparent that these latter are
at least two times overstated. Notice that when  the parameter in the
momentum distribution $\phi(|\vec k|)$, see Eq. (25), $\mu=100\ MeV$ the
branching ratios decreases three times (the $\mu^3$ low is natural for a
three-dimensional integral if a integrand has a range of the order of $\mu$).

   The imaginary part of the $\phi\rightarrow\gamma f_0(a_0)$ amplitude gives
90\% of the branching ratios  in the $\pi\pi (\pi\eta)$ channel, see also
Figs. 4 and 5.

   The real part of the $\phi\rightarrow\gamma f_0(a_0)$ amplitude dominates
in the  $K\bar K$ channel branching ratios , see also Fig.6 (a contribution of
the imaginary part of the amplitude in the $K^0\bar K^0$ channel is
negligible).

   During years of time it was believed by some that the decay $\phi\rightarrow
\gamma K_SK_S$ might prove an obstacle to research on the violation of CP
invariance in the decay $\phi\rightarrow K_LK_S$ despite the fact that it was
shown even in 1987 \cite{achasov-89} that $BR(\phi\rightarrow\gamma (f_0+a_0)
\rightarrow\gamma K_SK_S)\simeq 6.5\cdot 10^{-9}$ in the case which is the
worst case from the standpoint of research on the violation of CP invariance,
i.e., the case in which the $f_0(980)$ and $a_0(980)$ resonances are of a
four-quark nature ($q^2\bar q^2$). An upper limit in a Pickwickian sense
$BR(\phi\rightarrow\gamma (f_0+a_0)\rightarrow\gamma K_SK_S)\simeq
3.6\cdot 10^{-7}$ was given in \cite{achasov-92}. The following investigations,
see, for example, \cite{lucio-95}, only confirm these results. Here, see Table
III, we also confirm that the branching ratio for the decay
$\phi\rightarrow\gamma K_SK_S$ is low.

  The phase of $g_R(m)$ is the very important characteristic of the model
production because it causes the interference patterns in the reaction
$e^+e^-\rightarrow\gamma \pi^+\pi^-$ in the $\phi$ meson mass region. Emphasize
that just these interference patterns are used to identify the
$\phi\rightarrow\gamma f_0$ decay. As seen from Fig.3 the phases in the
molecule case (the solid curve) and in the four-quark (point-like) case
\cite{achasov-89} (the dashed curve) differ considerably. The calculation of
the interference patterns under discussion is a rather complex problem that
is a valid one for further investigation.

\acknowledgements

   N.N. Achasov thanks F.E. Close for many E-mail discussions and for copies of
personal notes, mailed kindly, and E.P. Solodov for many discussions of
interference patterns in the reaction $e^+e^-\rightarrow\gamma \pi^+\pi^-$
in the $\phi$ meson mass region. V.V. Gubin thanks International Soros
Science Education Program, Grant 962-S, for a financial support. V.I.
Shevchenko thanks the National Science Foundation and Russian Government,
Grant  NJ 77100, for a financial support.

\begin{figure}
\caption{ Diagrams for model. We send you immediatly by request to 
gubin@math.nsk.su in *.bmp format.} 
\end{figure}
\begin{figure}
\caption{ Branching ratio $BR(\phi\rightarrow\gamma R;m)$. The doted line
is the real part contribution. The dashed line is imaginary part contribution.
The solid line is the full branching ratio.} 
\end{figure}
\begin{figure}
\caption{ The phase $\delta=\arctan(Img_R(m)/Reg_R(m))$. The solid line
presents the m-dependence in the $K\bar K$ molecular case and the dashed one in
the $q^2\bar q^2$ case.}  
\end{figure}
\begin{figure}
\caption{ Mass spectra of $\pi\pi$ channel. The solid line is the full
spectra. The dashed and doted lines are the imaginary and real part 
contributions respectively. a) $\Gamma_0=50\ MeV$. The full branching ratio is
$1.9\cdot10^{-5}$. The branching ratio with account only imaginary part is
$1.7\cdot10^{-5}$. b) $\Gamma_0=100\ MeV$. The full branching ratio is
$2.2\cdot10^{-5}$. The branching ratio with account only imaginary part is
$2.0\cdot10^{-5}$.}         
\end{figure}
\begin{figure}
\caption{ Mass spectra of $\pi\eta$ channel. The solid line is the full
spectra. The dashed and doted lines are the imaginary and real part 
contributions respectively. a) $\Gamma_0=50\ MeV$. The full branching ratio is
$1.9\cdot10^{-5}$. The branching ratio with account only imaginary part is
$1.6\cdot10^{-5}$. b) $\Gamma_0=100\ MeV$. The full branching ratio is
$2.0\cdot10^{-5}$. The branching ratio with account only imaginary part is
$1.75\cdot10^{-5}$.}          
\end{figure}
\begin{figure}
\caption{ Mass spectra of $K^+K^-$ channel. $\Gamma_0=50\ MeV$.
The solid line is the full spectra. The dashed and doted lines are the 
imaginary and real part contributions respectively. The full branching ratio is
$1.0\cdot10^{-6}$. The branching ratio with account only imaginary part is
$0.9\cdot10^{-6}$.}             
\end{figure}
\begin{figure}
\caption{ Mass spectra of $K^0\bar K^0$ channel. $\Gamma_{a_0}=100\
MeV$ and $\Gamma_{f_0}=50\ MeV$ (see Table III). The ontribution of imaginary
part is negligible and not presented here. The branching ratio is
$2\cdot10^{-9}$.}                  
\end{figure}

\begin{table}
\caption{$BR(\phi\rightarrow\gamma a_0(f_0)\rightarrow\gamma\pi\eta
(\pi\pi))$}
\vspace*{1cm}
\begin{tabular}{|c|c|c|c|c|c|}
\multicolumn{6}{|c|}{$BR*10^5$} \\ \hline
\multicolumn{3}{|c|}{$\phi\rightarrow\gamma a_0\rightarrow\gamma\pi\eta$}
&\multicolumn{3}{c|}{$\phi\rightarrow\gamma f_0\rightarrow\gamma\pi
\pi$} \\ \hline
$m_{a_0};\>$GeV&$\Gamma_{a_0}=50$;\ MeV& $\Gamma_{a_0}=100$;\ MeV&
$m_{f_0};\>$GeV&$\Gamma_{f_0}=50$;\ MeV& $\Gamma_{f_0}=100$;\ MeV \\ \hline
0.960 & 2.232 & 2.298 & 0.960 & 2.357 & 2.564 \\ \hline
0.965 & 2.104 & 2.176 & 0.965 & 2.226 & 2.435 \\ \hline
0.970 & 1.959 & 2.045 & 0.970 & 2.077 & 2.297 \\ \hline
0.975 & 1.788 & 1.904 & 0.975 & 1.902 & 2.149 \\ \hline
0.980 & 1.575 & 1.753 & 0.980 & 1.685 & 1.991 \\ \hline
0.985 & 1.292 & 1.590 & 0.985 & 1.396 & 1.822 \\ \hline
0.990 & 0.919  & 1.049 & 0.990 &1.017 & 1.224 \\ \hline
0.995 & 0.697  & 0.902  & 0.995 & 0.792  & 1.072 \\
\end{tabular}
\end{table}
\begin{table}
\caption{ $ BR(\phi\rightarrow\gamma K^+K^-)$ }
\vspace*{1cm}
\begin{tabular}{|c|c|c|c|c|}
$m_{f_0}$;\ GeV  & $m_{a_0}$;\ GeV &$\Gamma_{f_0}$;\ MeV& $\Gamma_{a_0}$;
\ MeV &$BR*10^6$ \\ \hline

0.980 & 0.980  &50  &50 &1.01  \\ \hline
0.980 & 0.980  &100 &100&0.48   \\ \hline
0.980 & 0.980  &50  &100 &0.715  \\ \hline
0.990 & 0.990  &100 &100 &0.6    \\ \hline
0.980 & 0.990  &50  &100 &0.72   \\ \hline
0.990 & 0.980  &50  &100 &1.0    \\ \hline
0.990 & 0.990  &50  &100 &1.12  \\ \hline
0.990 & 0.990  &50  &50  &1.81 \\
\end{tabular}
\end{table}
\begin{table}
\caption{ $ BR(\phi\rightarrow\gamma K^0\bar K^0)$}
\vspace*{1cm}
 \begin{tabular}{|c|c|c|c|c|}
 $m_{f_0}$;\ GeV  & $m_{a_0}$;\ GeV &$\Gamma_{f_0}$;\ MeV& $\Gamma_{a_0}$;
\ MeV &$BR*10^9$ \\ \hline

0.980 & 0.980  &50  &50 & $1.57\cdot 10^{-3}$  \\ \hline
0.980 & 0.980  &100 &100& $2.127\cdot 19^{-3}$   \\ \hline
0.980 & 0.980  &50  &100 &2  \\ \hline
0.990 & 0.990  &100 &100 &$9.48\cdot 10^{-4}$  \\ \hline
0.980 & 0.990  &50  &100 &5.1   \\ \hline
0.990 & 0.980  &50  &100 &6.5    \\ \hline
0.990 & 0.990  &50  &100 &4.3  \\ \hline
0.990 & 0.990  &50  &50  &$1.1\cdot 10^{-3}$ \\
\end{tabular}
\end{table}


\begin{references}
\bibitem{achasov-84}
N.N. Achasov, S.A. Devyanin and G.N. Shestakov, Usp. Fiz. Nauk 142 (1984) 361.
\bibitem{achasov-91}
N.N. Achasov, Nucl. Phys. B (Proc. Suppl.) 21 (1991) 189.
\bibitem{achasov-1991}
N.N. Achasov and G.N. Shestakov, Usp. Fiz. Nauk 161 (1991) 53.
\bibitem{jaffe-77}
R.L. Jaffe, Phys. Rev. D 15 (1977) 267, 281.
\bibitem{weinstein-90}
J. Weinstein and N. Isgur, Phys. Rev. D 41 (1990) 2236.
\bibitem{close-93}
F.E. Close, Yu.L. Dokshitzer, V.N. Gribov, V.A. Khoze and M.G. Ryskin,
Phys. Lett. B 319 (1993) 291.
\bibitem{achasov-89}
N.N. Achasov and V.N.Ivanchenko, Nucl. Phys. B  315 (1989) 465,\\
Preprint INP 87-129 (1987), Novosibirsk.
\bibitem{isgur-93}
F.E. Close, N. Isgur and S. Kumano, Nucl. Phys. B 389 (1993) 513.
\bibitem{particle-94}
Particle Data Group, Phys. Rev. D 50 (1994) 1468, 1469.
\bibitem{achasov-95}
N.N. Achasov, THE SECOND DA$\Phi$NE PHYSICS HANDBOOK, Vol. II, edited by
L. Maiani, G. Pancheri, N. Paver, dei Laboratory Nazionali di Frascati,
Frascati, Italy ( May 1995), p. 671.\\
N.N. Achasov and V.V. Gubin, Pis'ma v ZhETF  62 (1995) 182.\\
N.N. Achasov and V.V. Gubin, Phys. Lett. B 363 (1995) 106.
\bibitem{achasov-88}
N.N. Achasov and G.N. Shestakov, Z. Phys. C  41 (1988) 309.
\bibitem{close-92}
N. Brown and F.E. Close, THE DA$\Phi$NE PHYSICS HANDBOOK, Vol. II, edited by
L. Maiani, G. Pancheri, N. Paver, dei Laboratory Nazionali di Frascati,
Frascati, Italy ( December 1992), p. 447.
\bibitem{achasov-92}
N.N. Achasov, Pis'ma v ZhETF  55 (1992) 373.
\bibitem{lucio-95}
J.L. Lucio, M. Napsuciale, Nucl. Phys. B 440 (1995) 237.
\end{references}
\end{document}